\documentclass[12pt]{article}
\usepackage{amssymb,amsmath,epsfig}

\begin{document}

\title{\bf Inflationary Weak Anisotropic Model with General Dissipation Coefficient}
\author{M. Sharif \thanks {msharif.math@pu.edu.pk} and Rabia Saleem
\thanks{rabiasaleem1988@yahoo.com}\\
Department of Mathematics, University of the Punjab,\\
Quaid-e-Azam Campus, Lahore-54590, Pakistan.}

\date{}
\maketitle

\begin{abstract}
This paper explores the dynamics of warm intermediate and
logamediate inflationary models during weak dissipative regime with
a general form of dissipative coefficient. We analyze these models
within the framework of locally rotationally symmetric Bianchi type
I universe. In both cases, we evaluate solution of inflaton,
effective scalar potential, dissipative coefficient, slow-roll
parameters, scalar and tensor power spectra, scalar spectral index
and tensor to scalar ratio under slow-roll approximation. We
constrain the model parameters using recent data and conclude that
anisotropic inflationary universe model with generalized dissipation
coefficient remains compatible with WMAP9, Planck and BICEP2 data.
\end{abstract}

\section{Introduction}

The standard universe model successfully explains the observations
of CMBR but there are still some unresolved issues. Inflationary
cosmology is proved to be a cornerstone to resolve long-standing
theoretical issues including horizon problem, flatness, magnetic
monopole issue and origin of fluctuations. Scalar field as a primary
ingredient of inflation provides the causal interpretation of the
origin of LSS distribution and observed anisotropy of CMB \cite{6}.
Inflationary standard models are classified into slow-roll and
reheating epochs. In slow-roll period, potential energy dominates
kinetic energy and all interactions between scalar (inflatons) and
other fields are neglected, hence the universe inflates \cite{7}.
Subsequently, the universe enters into reheating period where the
kinetic energy is comparable to potential energy. Thus the inflaton
starts an oscillation about minimum of its potential losing their
energy to other fields that present in the theory \cite{8}. After
this epoch, the universe is filled with radiation.

Warm inflation \cite{B} (opposite to cold inflation) has an
attractive feature of joining the end stage of inflation with the
current universe. It is distinguished from cold inflation in a way
that thermal radiation production occurs during inflationary epoch
and reheating period is avoided. The dissipation effects become
strong enough due to the production of thermal fluctuations of
constant density which play a vital role in the formation of initial
fluctuations necessary for LSS formation. The thermalized particles
are produced continuously by microscopic processes which must occur
at a timescale much faster than Hubble scale $(H)$, hence the decay
rates of the particles must be larger than $H$. During this regime,
density fluctuation arises from thermal rather than quantum
fluctuation \cite{9}. Warm inflationary era ends when the universe
stops inflating, then the universe enters into the radiation
dominated phase smoothly. Finally, the remaining inflatons or
dominant radiation fields create matter components of the universe
\cite{10}. The feasibility of the warm inflation scenario from
various view points is discussed in \cite{A}. Their results as a
whole show that it is extremely difficult to realize the idea of
warm inflation.

Dissipative effects could lead to a friction term in the equation of
motion for an inflaton field during the inflationary era. The
friction term may be linear as well as localized and is described by
a dissipation coefficient. Bastero-Gil et al. \cite{11} made
considerable explicit calculations using quantum field theory method
that compute all the relevant decay and scattering rates in the warm
inflationary models. Berera et al. \cite{12} presented particular
scenario of low-temperature regimes in the context of dissipation
coefficient. They considered the value of dissipation coefficient in
supersymmetric (SUSY) models which have an inflaton together with
multiplets of heavy and light fields. Dissipation leads to two
important consequences: weak $(\Gamma\ll H)$ and strong regimes
$(\Gamma\gg H)$. During weak dissipative regime, the primordial
density perturbation spectrum is determined by thermal fluctuations
rather than vacuum fluctuations while restrictions on the gradient
of inflaton potential may be relaxed in strong regime.

Inflationary universe has many interesting exact solutions that can
be found using an exponential potential often called a power-law
inflation. Here, the scale factor has a power-law type evolution
$a(t)=t^{p}$, where $p>1$ \cite{14}. Another exact solution is
obtained in the de Sitter inflationary universe where a constant
scalar potential is considered \cite{15}. Exact solutions of the
inflationary cosmology can also be obtained for two particular
scenarios, i.e., intermediate and logamediate with specific growth
of the scale factors \cite{16,17}. This type of expansion is slower
than de Sitter inflation but faster than power-law inflation, so
dubbed as ``intermediate". The string or M theory motivates
intermediate inflation according to which higher order curvature
invariant corrections to the Einstein-Hilbert action must be
proportional to Gauss-Bonnet terms for the ghost free action. These
terms arise naturally as the leading order in the expansion of
inverse string tension $``\alpha"$ to low energy string effective
action. It has been found that 4-dimensional Gauss-Bonnet
interaction with dynamical dilatonic scalar coupling leads to a
solution, i.e., intermediate form of the scale factor \cite{18}. On
the other hand, logamediate inflation (generalized model of the
expanding universe) is motivated by applying weak general conditions
on the indefinite expanding cosmological models \cite{19}.

These models were originally developed as exact solutions of
inflationary cosmology but were best formulated using slow-roll
approximation. During slow-roll approximation, it is possible to
find a spectral index $n_{s}=1$. In particular, intermediate
inflation leads to $n_{s}=1$ for special value $f^{\ast}=2/3$ (but
this value is not supported by the current observational data
\cite{6}) that corresponds to the Harrizon-Zel'dovich spectrum
\cite{20}. In both models, an important observational quantity is
the tensor to scalar ratio $(r)$, which is significantly non-zero
\cite{21}. Recently, the effects from BICEP2 experiment of
gravitational waves in the B-mode have been analyzed which predict
$r={0.2}_{-0.05}^{+0.07}$ $(68\%$ C.L.) and take out the value $r=0$
at a significance of $7.0\sigma$ \cite{21a}. Therefore, the tensor
modes should not be neglected.

Setare and Kamali \cite{22} studied ``warm inflation" with vector
and non-abelian gauge fields during intermediate and logamediate
scenarios using flat FRW background with constant and variable
dissipation coefficients. Motivated by this, we have discussed
inflation with both types of field and proved that locally
rotationally symmetric (LRS) Bianchi I (BI) universe model is also
compatible with WMAP7 observations \cite{23}. del Campo and Herrera
\cite{24} explored dynamics of warm-Chaplygin inflationary universe
model and discussed cosmological perturbations in warm inflationary
universe model with viscous pressure. Herrera et al. \cite{25}
studied intermediate generalized Chaplygin gas (CG) inflationary
universe model with standard as well as tachyon scalar fields and
checked its compatibility with WMAP7 data. We have studied
inflationary dynamics of generalized cosmic CG using standard and
tachyon scalar fields (with and with out viscous pressure) during
intermediate and logamediate scenarios \cite{26}. Setare and Kamali
\cite{27} investigated dynamics of warm inflation with viscous
pressure in FRW universe model and on the brane with constant as
well as variable dissipation and bulk viscosity coefficients. We
have extended this work to LRS BI universe model \cite{28}.

Herrera et al. \cite{29} analyzed the possible realization of an
expanding intermediate and logamediate scale factors within the
framework of a warm inflationary FRW as well as loop quantum
cosmology models. They checked that how both types of inflation work
with a generalized form of dissipative coefficient during weak and
strong dissipative regimes. The FRW universe is just an
approximation to the universe we see as it ignores all the structure
and other observed anisotropies, e.g., in the CMB temperature
\cite{30}. One of the great achievements of inflation is having a
naturally embedded mechanism to account for these anisotropies.
Although the new era of high precision cosmology of CMB radiation
improves our knowledge to understand the infant as well as the
present day universe. There arises a question about the main
assumption of an exact isotropy of the CMB. There are two pieces of
observational evidence demonstrating that there is no exact
isotropy. The first is the existence of small anisotropy deviations
from isotropy of the CMB radiation and second the presence of large
angle anomalies that are shown as real features by the Planck
satellite results. This helps to construct an alternative model to
decode effects of the early universe on the present day LSS without
affecting the processes of nucleosythesis \cite{58}.

Bianchi models can be alternatives to the standard FRW models with
small deviations from exact isotropy to explain anisotropies and
anomalies in the CMB. Martinez-Gonzalez and Sanz \cite{59} proved
that the small quadrupole component of CMB temperature found by COBE
implies that if the universe is homogeneous but anisotropic BI then
there must be a small departure from the flat Friedmann model.
Bianchi spacetimes are geometries with spatially homogeneous
(constant $t$) surfaces which are invariant under the action of a
three dimensional Lie group. Bianchi type I, the straight forward
generalization of the flat FRW metric with symmetry group described
by $C^{k}$$_{ij}=0$ corresponds to flat hypersurfaces. Following
this idea, we extend this work to LRS BI universe model which is a
generalization of our previous paper \cite{23} in which a particular
choice of dissipation coefficient $(n=3)$ is considered.

The paper is organized as follows. Section \textbf{2} provides basic
formalism of warm inflation for LRS BI universe model. In section
\textbf{3}, we deal with weak dissipative regime and develop the
model in two particular scenarios (i) intermediate inflation (ii)
logamediate inflation. We evaluate explicit expressions for
inflaton, potential and rate of decay as well as perturbation
parameters. The behavior of these physical parameters is checked
through graphical analysis by constraining the model parameters with
recent observations. Finally, the results are summarized in section
\textbf{4}.

\section{Anisotropic Warm Inflationary Model}

In this section, we present basic formalism of warm inflation in the
background of LRS BI universe model whose line element is given as
\cite{30}
\begin{equation*}
ds^2=-dt^2+X^2(t)dx^2+Y^2(t)(dy^2+dz^2),
\end{equation*}
where $X(t)$ and $Y(t)$ denote the expansion measure along $x$-axis
and $y,~z$-axis, respectively. Under a linear relationship,
$X=Y^{\mu}$ ($\mu\neq1$ be the anisotropic parameter), the above
metric is reduced to
\begin{equation*}
ds^2=-dt^2+Y^{2\mu}(t)dx^2+Y^2(t)(dy^2+dz^2).
\end{equation*}
The basic ingredients of the universe are assumed to be
self-interacting scalar field ($\psi$) and radiation field
$(\gamma)$. The inflaton possesses following energy density
$(\rho_{\psi})$ and pressure $(P_{\psi})$, respectively
\begin{equation}\label{1}
\rho_{\psi}=\frac{\dot{\psi}^2}{2}+V(\psi),\quad
P_{\psi}=\frac{\dot{\psi}^2}{2}-V(\psi),
\end{equation}
where $V(\psi)$ is the effective potential associated with $\psi$
and dot stands for derivative with respect to cosmic time $t$. The
anisotropic warm inflation is described by the first evolution
equation and conservation equations of inflaton and radiation given
by
\begin{eqnarray}\nonumber
H^2_{2}=\frac{\kappa}{1+2\mu}(\rho_{\psi}+\rho_{\gamma})
&=&\frac{\kappa}{1+2\mu}\left(\frac{\dot{\psi}^2}{2}+V(\psi)+\rho_{\gamma}\right),
\\\nonumber\dot{\rho_{\psi}}+(\mu+2)H_{2}(\rho_{\psi}+P_{\psi})
&=&-\Gamma\dot{\psi}^{2},\\\label{2}
\dot{\rho}_{\gamma}+\frac{4}{3}(\mu+2)H_{2}\rho_{\gamma}
&=&\Gamma\dot{\psi}^{2},
\end{eqnarray}
where $\rho_{\gamma}$ is the radiation density, $H_{2}$ is the
directional Hubble parameter and dissipation factor $(\Gamma)$ is
introduced to measure the decay rate. The second law of
thermodynamics suggests $\Gamma>0$ implying that $\rho_{\psi}$
dissipates into $\rho_{\gamma}$. It is found that it can be
considered as a constant, function of inflaton $(\Gamma(\psi))$,
function of temperature $(\Gamma(T))$, function of both
$(\Gamma(\psi,T))$ and equivalent to $V(\psi)$ in some papers
\cite{9}.

Here, we take a general form of the dissipative coefficient as
\begin{equation}\label{3}
\Gamma=C_{\psi}\frac{T^{n}}{\psi^{n-1}},
\end{equation}
where $n$ be any arbitrary integer and $C_{\psi}$ is associated to
the dissipative microscopic dynamics \cite{31}. In this reference,
Zhang and Basero-Gil et al. analyzed different choices of the
integer $n$ which correspond to different expressions for
dissipation coefficient. In particular, for $n=3$,
$C_{\psi}=0.64h^{4}\mathcal{N}$, where
$\mathcal{N}=\mathcal{N}_{\chi}\mathcal{N}^{2}_{decay}$
($\mathcal{N}_{\chi}$ is the multiplicity of the $\mathcal{X}$
superfield and $\mathcal{N}$ is the number of decay channels
available in $\mathcal{X}$'s decay) \cite{12,31,32}. The value $n=1$
leads to $\Gamma\propto T$ (represents the high-temperature SUSY
case), $n=0$ generates $\Gamma\propto\psi$ (corresponds to an
exponentially decaying propagator in the SUSY case) and $n=-1$ leads
the decay rate $\Gamma\propto\frac{\psi^{2}}{T}$ (corresponds to the
non-SUSY case). The case $n=3$ implies the most common form
$\Gamma\sim \frac{T^3}{\phi^2}$ considered for the warm intermediate
and logamediate models \cite{33}. For convenience, we only focus on
the parameter regime $|n|<4$. In particular, we have paid attention
to the cosmological results, i.e., the potential takes the monomial
form and the hybrid-like form when $n=-1,~0,~1$.

During inflationary regime, stable regime can be obtained by
applying an approximation, i.e., $\rho_{\psi}\approx
V(\psi),~\rho_{\psi}>\rho_{\gamma}$. Under this limit, the evolution
equation is reduced to
\begin{equation}\label{4}
H^2_{2}=\frac{\kappa}{1+2\mu}\rho_{\psi}=\frac{\kappa}{1+2\mu}V(\psi).
\end{equation}
Using this equation along with conservation equation of inflaton, we
have
\begin{equation}\label{5}
\dot{\psi}^{2}=\frac{2(1+2\mu)(-\dot{H_{2}})}{(\mu+2)\kappa(1+R)},
\end{equation}
where decay rate is the ratio of $\Gamma$ by $H_{2}$, given by
$R=\frac{\Gamma}{(\mu+2)H_{2}}$. In warm inflation, the radiation
production is assumed to be quasi-stable where
$\dot{\rho_{\gamma}}\ll\frac{4}{3}(\mu+2)H_{2}\rho_{\gamma},
~\dot{\rho}_{\gamma}\ll\Gamma\dot{\psi}^2$. Using Eq.(\ref{5}) and
quasi-stable condition in the last equation of Eq.(\ref{2}), it
follows that
\begin{equation}\label{6}
\rho_{\gamma}=\frac{3(1+2\mu)\Gamma(-\dot{H_{2}})}{2\kappa
(\mu+2)^2(1+R)H_{2}}=C_{\gamma}T^4,
\end{equation}
where $C_{\gamma}=\frac{\pi^2g_{\ast}}{30}$ in which $g_{\ast}$ is
known as the number of relativistic degrees of freedom. The
temperature of thermal bath can be extracted from second equality of
the above equation as
\begin{equation}\label{7}
T=\left[\frac{3(1+2\mu)\Gamma(-\dot{H_{2}})}{2\kappa C_{\gamma}
(\mu+2)^2(1+R)H_{2}}\right]^{\frac{1}{4}}.
\end{equation}
Substituting the value of $T$ in Eq.(\ref{3}), we have
\begin{equation}\label{8}
\Gamma^{\frac{4-n}{4}}=\alpha_{n}(1+R)^{-\frac{n}{4}}
\left(\frac{-\dot{H_{2}}}{H_{2}}\right)^{\frac{n}{4}}\psi^{1-n},
\end{equation}
where $\alpha_{n}=C_{\psi}\left[\frac{3(1+2\mu)}{2\kappa
C_{\gamma}(\mu+2)^2}\right]^{\frac{n}{4}}$. The effective potential
can be obtained from first evolution equation with the help of
Eqs.(\ref{5}) and (\ref{6}) as
\begin{equation}\label{9}
V(\psi)=\left(\frac{1+2\mu}{\kappa}\right)H_{2}^{2}+\frac{(1+2\mu)\dot{H_{2}}}
{(\mu+2)\kappa(1+R)}\left[1+\frac{3}{2}R\right].
\end{equation}
In the following, we shall develop warm inflationary model during
intermediate and logamediate scenarios for weak dissipative regime.

\section{Weak Dissipative Regime}

Here, we consider that warm anisotropic inflationary model evolves
according to weak dissipative regime where $\Gamma\ll(\mu+2)H_{2}$.

\subsection{Intermediate Era}

In this era, the scale factor evolution is given by \cite{16}
\begin{equation}\label{10}
Y(t)=Y_{0}\exp(A^{\ast}t^{f^{\ast}}),\quad A^{\ast}>0,~0<f^{\ast}<1.
\end{equation}
Using this scale factor in Eq.(\ref{5}), we find solution of
inflaton field as
\begin{equation}\label{11}
\psi(t)-\psi(t_{0})=\alpha_{0}t^{\frac{f^{\ast}}{2}},
\end{equation}
where
$\alpha_{0}=\frac{2}{f^{\ast}}\left[\frac{2(1+2\mu)}{\kappa(\mu+2)}(A^{\ast}f^{\ast})
(1-f^{\ast})\right]^{\frac{1}{2}}$. Substituting the value of $t$
from the above solution with $\psi(t_{0})=0$, we obtain $H_{2}$ in
terms of $\psi$ as
\begin{equation}\label{12}
H_{2}(\psi)=(A^{\ast}f^{\ast})t^{f^{\ast}-1}=(A^{\ast}f^{\ast})\left(\frac{\alpha_{0}}{\psi(t)}
\right)^{\frac{2(1-f^{\ast})}{f^{\ast}}}
\propto{\psi(t)}^{\frac{2(f^{\ast}-1)}{f^{\ast}}}.
\end{equation}
In weak dissipative regime, $V(\psi)$ turns out to be
\begin{equation}\nonumber
V(\psi)=\left(\frac{1+2\mu}{\kappa}\right)(A^{\ast}f^{\ast})^2
\left(\frac{\alpha_{0}}{\psi}\right)^{\frac{4(1-f^{\ast})}{f^{\ast}}}.
\end{equation}
The friction term under weak dissipation is reduced to
$\Gamma^{\frac{4-n}{4}}=\alpha_{n}
(\frac{-\dot{H_{2}}}{H_{2}})^{\frac{n}{4}}\psi^{1-n}$ and can be
written in terms of inflaton as
\begin{equation}\label{14}
\Gamma(\psi)=C_{\psi}^{\frac{4}{4-n}}\left[\frac{3(1+2\mu)(1-f^{\ast})
\alpha_{0}^{\frac{2}{f^{\ast}}}}{2\kappa
C_{\gamma}(\mu+2)^2}\right]^{\frac{n}{4-n}}
\psi^{\frac{4f^{\ast}(1+n)-2n}{f^{\ast}(4-n)}}.
\end{equation}

In \cite{11}, a detailed analysis of inflationary scenario with
anisotropy is proposed. Remarkably, they have found that degrees of
anisotropy are universally determined by the slow-roll parameter.
Since the slow-roll parameter is observationally known to be of the
order of a percent, therefore anisotropy during inflation cannot be
entirely negligible. In order to analyze the slow-roll dynamics, the
dimensionless slow-roll parameters are $\epsilon$ and $\eta$
\cite{34}, where $\epsilon$ is the standard slow-roll parameter
while $\eta$ can be expressed in terms of $\epsilon$. A sensible
inflation not only demands $\epsilon\ll1$ but also $\eta$ must be
small over a reasonably large time period. These parameters are
represented as a function of $\psi$
\begin{eqnarray}\nonumber
\epsilon&=&-\left(\frac{3}{\mu+2}\right)\frac{\dot{H}_{2}}{H^{2}_{2}}
=\left(\frac{3}{\mu+2}\right)\left(\frac{1-f^{\ast}}{A^{\ast}f^{\ast}}\right)\alpha_{0}^{2}
\psi^{-2},\\\nonumber
\eta&=&-\left(\frac{3}{\mu+2}\right)\frac{\ddot{H}_{2}}{H_{2}\dot{H}_{2}}=
\left(\frac{3}{\mu+2}\right)\left(\frac{2-f^{\ast}}{A^{\ast}f^{\ast}}\right)\alpha_{0}^{2}
\psi^{-2}.
\end{eqnarray}
The condition for the occurrence of inflation $(\epsilon<1)$ is
satisfied when inflaton is restricted to
\begin{equation}\nonumber
\psi>\left[\left(\frac{3}{\mu+2}\right)\left(\frac{1-f^{\ast}}
{A^{\ast}f^{\ast}}\right)\right]^{\frac{1}{2}}\alpha_{0}.
\end{equation}
The initial inflaton $(\psi_{1})$ is obtained at the earliest
possible inflationary stage where $\epsilon=1$
\begin{equation}\nonumber
\psi_{1}=\left[\left(\frac{3}{\mu+2}\right)\left(\frac{1-f^{\ast}}
{A^{\ast}f^{\ast}}\right)\right]^{\frac{1}{2}}\alpha_{0}.
\end{equation}
The number of e-folds $(N)$ interpolated between two different times
$t_{1}$ (beginning of inflation) and $t_{2}$ (end of inflation) is
defined as follows
\begin{equation}\label{15}
N=\left(\frac{\mu+2}{3}\right)\int^{t_{2}}_{t_{1}}H_{2}dt=\left(\frac{\mu+2}{3}\right)
A^{\ast}\alpha_{0}^{-2}\left(\psi_{2}^{2}-\psi_{1}^{2}\right).
\end{equation}
The second equality shows $N$ in terms of $\psi$ using
Eq.(\ref{12}).

Perturbations are usually characterized in terms of four quantities,
i.e., scalar (tensor) power spectra $(P_{R}(k),P_{T}(k))$ ($k$ be
the wave number) and corresponding scalar (tensor) spectral indices
$(n_{s},n_{T})$. The power spectrum is introduced to measure the
variance in the fluctuations produced by inflaton \cite{35}. For a
standard scalar field, the density perturbation could be written as
$P_{R}^{\frac{1}{2}}=\left(\frac{\mu+3}{2}\right)\frac{H_{2}}
{\dot{\psi}}\delta\psi$ \cite{9}. In weak dissipative regime,
$\delta\psi^2=\left(\frac{\mu+3}{2}\right)H_{2}T$ \cite{10,10a}. The
scalar power spectrum is obtained by combining Eqs.(\ref{7}),
(\ref{12}) and (\ref{14}) as
\begin{eqnarray}\nonumber
P_{R}(k)=\left(\frac{\mu+2}{3}\right)^{3}\frac{T}{\psi^{2}}H_{2}^{3}
&=&\left(\frac{\kappa(\mu+2)^{4}}{6(1+2\mu)}\right)
\left[\frac{3(1+2\mu)C_{\psi}}{2\kappa
C_{\gamma}(\mu+2)^2}\right]^{\frac{1}{4-n}}H_{2}^{\frac{11-3n}{4-n}}
\\\label{16}&\times&(-\dot{H_{2}})^{\frac{n-3}{4-n}}
\psi^{\frac{1-n}{4-n}}=\alpha_{1}\psi^{-\beta_{1}},
\end{eqnarray}
where
\begin{eqnarray}\nonumber
\alpha_{1}&=&\left(\frac{\kappa(\mu+2)^{4}}{6(1+2\mu)}\right)
\left[\frac{3(1+2\mu)C_{\psi}}{2\kappa
C_{\gamma}(\mu+2)^2}\right]^{\frac{1}{4-n}}(A^{\ast}f^{\ast})^{2}
(1-f^{\ast})^{-\frac{3-n}{4-n}}\alpha_{0}^{\frac{1-n}{4-n}+\beta_{1}},\\\nonumber
\beta_{1}&=&\frac{10-2n-f^{\ast}(17-5n)}{f^{\ast}(4-n)}.
\end{eqnarray}
The scalar spectral index for our model is defined as
\begin{equation}\nonumber
n_{s}-1=\frac{d\ln P_{R}(k)}{d\ln
k}=\left[\frac{12(1+2\mu)(f^{\ast}-1)}{(\mu+2)^{2}\kappa
f^{\ast}}\right]\left[\frac{10-2n-f^{\ast}(17-5n)}
{f^{\ast}(4-n)}\right]\psi^{-2}.
\end{equation}
Inserting $\psi_{1}$ in the expression of $N$, we obtain the value
of final inflaton
\begin{equation}\nonumber
\psi_{2}=\left[\alpha_{0}^{2}\left(\frac{3}{\mu+2}\right)\left(\frac{N}{A^{\ast}}
+\frac{1-f^{\ast}}{A^{\ast}f^{\ast}}\right)\right]^{\frac{1}{2}}.
\end{equation}
Using this equation, $n_{s}$ becomes
\begin{equation}\label{e}
n_{s}=1-\left[\frac{10-2n-f^{\ast}(17-5n)}{2(4-n)(1+f^{\ast}(N-1))}\right],
\end{equation}
which yields the specific value of $f^{\ast}$ in terms of $N$ and
$n_{s}$
\begin{equation}\label{a}
f^{\ast}=1-\frac{10-2n-2(4-n)(1-n_{s})}{7-3n+2(4-n)(N-1)(1-n_{s})}.
\end{equation}
The parameter $A^{\ast}$ can be calculated through Eq.(\ref{16}) as
\begin{eqnarray}\nonumber
A^{\ast}&=&\alpha_{2}\left[\frac{2(1+2\mu)C_{\psi}}{2\kappa
C_{\gamma}(\mu+2)^{2}}\right]^{\frac{f^{\ast}}{2n-10+2f^{\ast}(13-4n)}}(1+f^{\ast}(N-1))
^{\frac{10-2n-f^{\ast}(17-5n)}{2(2n-10+f^{\ast}(13-4n))}}\\\label{b}&\times&P_{R}
^{\frac{f^{\ast}(4-n)}{2n-10+f^{\ast}(13-4n)}},
\end{eqnarray}
where
\begin{eqnarray}\nonumber
\alpha_{2}&=&\left(\frac{\kappa(\mu+2)^{4}}{6(1+2\mu)}\right)
^{\frac{f^{\ast}(4-n)}{2n-10+2f^{\ast}(13-4n)}}
\left(\frac{8(1+2\mu)(1-f^{\ast})}{\kappa(\mu+2)}\right)
^{\frac{2n-10+f^{\ast}(19-7n)}{2(2n-10+2f^{\ast}(13-4n))}}\\\nonumber
&\times&\left(\frac{\mu+2}{3}\right)^{\frac{10-2n-f^{\ast}(17-5n)}
{2(2n-10+2f^{\ast}(13-4n))}}{f^{\ast}}^{\frac{f^{\ast}(7-n)}{(4-n)(2n-10+2f^{\ast}(13-4n))}}.
\end{eqnarray}
The rate of decay as a function of $n_{s}$ can be calculated as
follows
\begin{eqnarray}\nonumber
R&=&\left[\frac{12(1+2\mu)(1-f^{\ast})(10-2n-f^{\ast}(17-5n))}{\kappa
(\mu+2)^{2}{f^{\ast}}^{2}(4-n)(1-n_{s})}\right]^{\frac{2(2-n)+f^{\ast}(3n-2)}{f^{\ast}(4-n)}}\\\label{c}
&\times&\frac{\kappa}{(\mu+2)(A^{\ast}
f^{\ast})\alpha_{0}^{\frac{2(1-f^{\ast})}{f^{\ast}}}}.
\end{eqnarray}
Figures \textbf{1} and \textbf{2} verify that $R<1$ for specific
values of the parameters.
\begin{figure}
\centering\epsfig{file=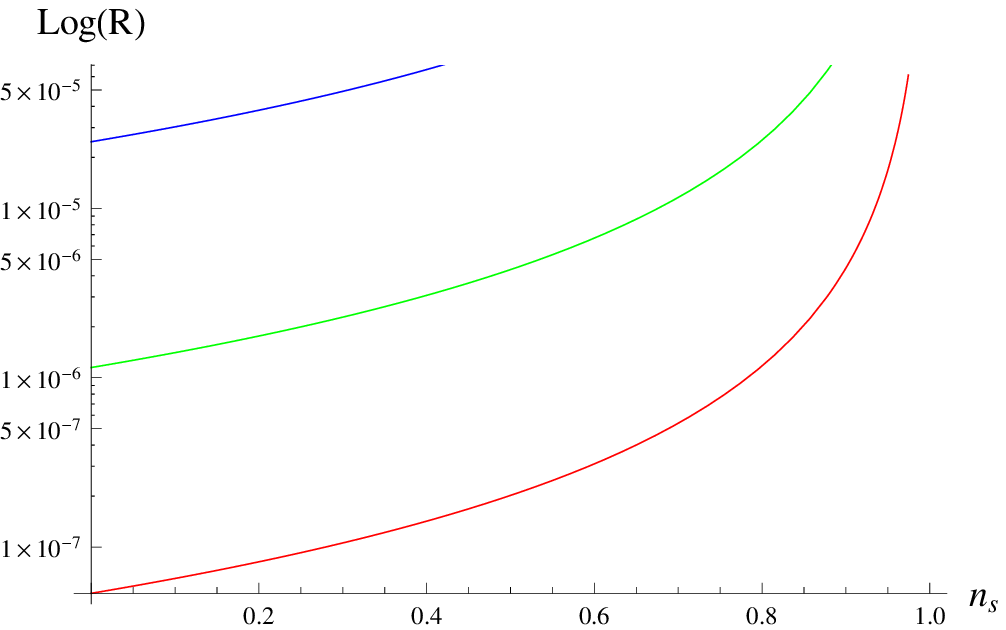,
width=0.50\linewidth}\epsfig{file=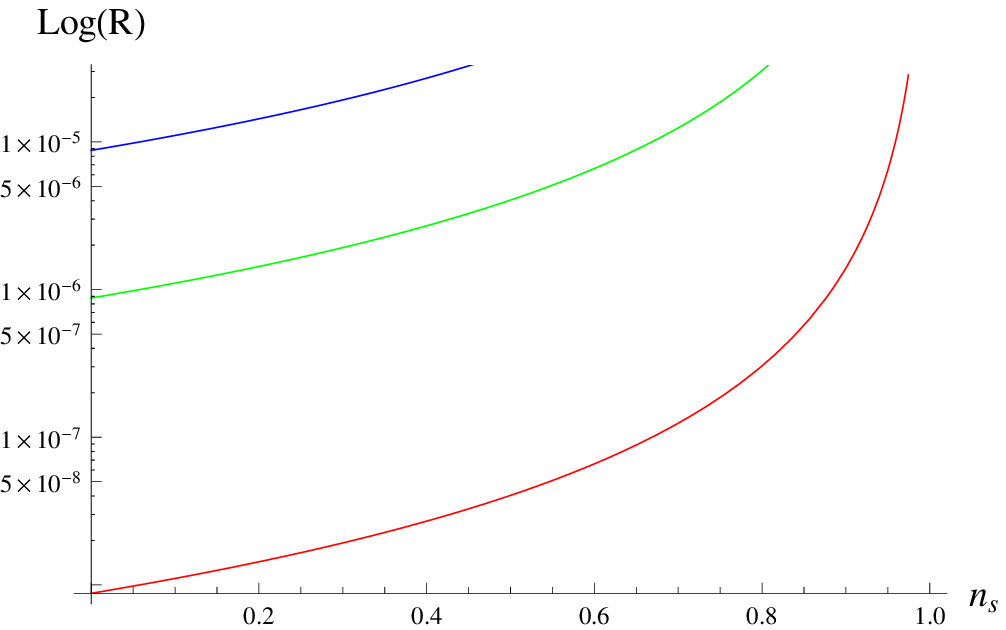,
width=0.50\linewidth}\caption{Variations in Log$(R)$ versus $n_{s}$:
Left plot for $f^{\ast}=0.42,~\mu\approx0.5,~n=1,~C_{\gamma}=70$,
$C_{\psi}=10^{-6}$ (red), $C_{\psi}=10^{-5}$ (green),
$C_{\psi}=10^{-4}$ (blue); ~~~(Right)
$f^{\ast}=0.37,~n=0,~C_{\psi}=10^{-9}$ (red), $C_{\psi}=10^{-7}$
(green), $C_{\psi}=10^{-6}$ (blue) during intermediate era.}
\end{figure}
\begin{figure}
\centering\epsfig{file=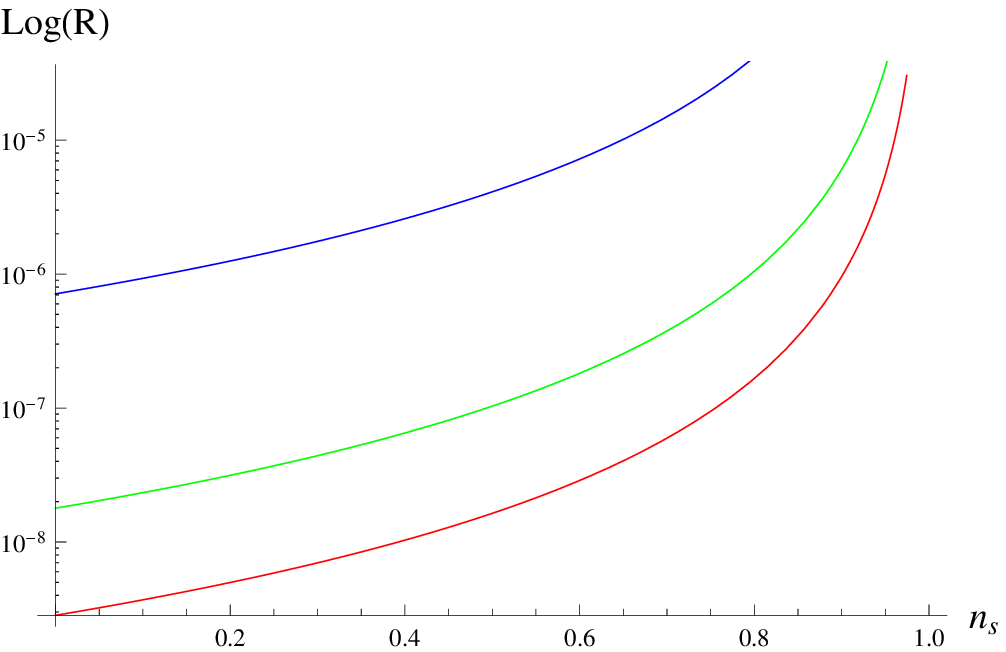,
width=0.50\linewidth}\epsfig{file=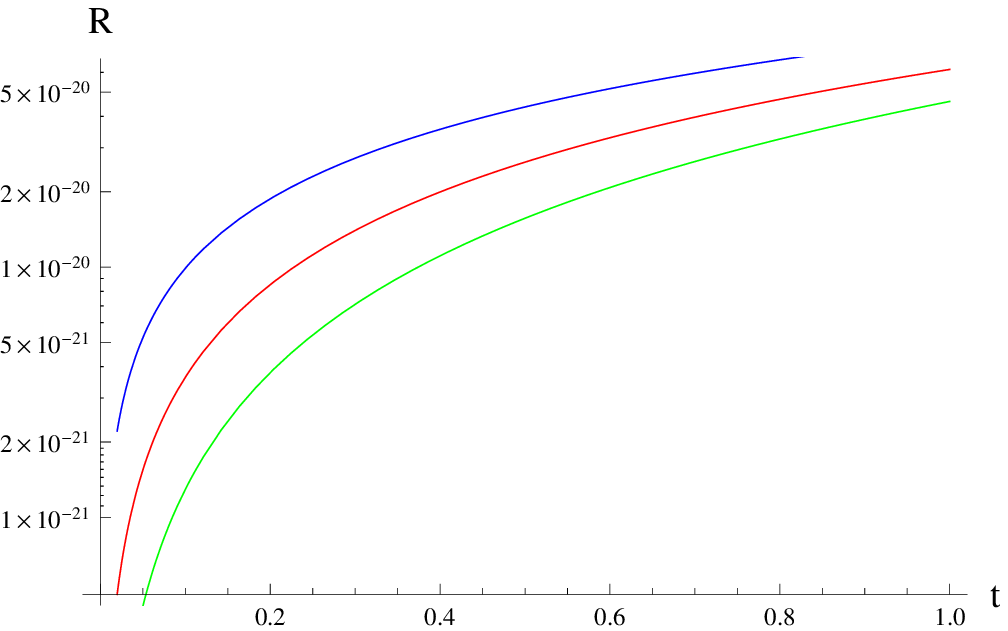,
width=0.50\linewidth}\caption{Log$(R)$ versus $n_{s}$ for
$f^{\ast}=0.34,~\mu\approx0.5,~n=-1,~C_{\gamma}=70$,
$C_{\psi}=10^{-12}$ (red), $C_{\psi}=10^{-11}$ (green),
$C_{\psi}=10^{-9}$ (blue) (Left). Variations in $R$ versus $t$ for
$n=0$ (red), $n=-1$ (green), $n=1$ (blue) (Right).}
\end{figure}

The corresponding tensor perturbations are
\begin{equation}\label{17a}
P_{T}(k)=\frac{2\kappa}{9\pi^2}(\mu+2)^2H_{2}^{2}=\frac{2\kappa}{9\pi^2}
(\mu+2)^2(A^{\ast}f^{\ast})^2\left(\frac{\psi}{\alpha_{0}}\right)
^{\frac{4(f^{\ast}-1)}{f^{\ast}}}.
\end{equation}
Combining Eqs.(\ref{16}) and (\ref{17a}), the tensor to scalar ratio
(in terms of $\psi,~N,~n_{s}$) becomes
\begin{eqnarray}\nonumber
r(k)&=&\alpha_{2}\psi^{\beta_{2}}=\alpha_{2}\left[\frac{24(1+2\mu)
(1-f^{\ast})(1+f^{\ast}(N-1))}{\kappa
(\mu+2)^{2}{f^{\ast}}^{2}}\right]\\\label{d}&=&\alpha_{2}
\left[\frac{12(1+2\mu)(1-f^{\ast})(10-2n-f^{\ast}(17-5n))}{\kappa
(\mu+2)^{2}{f^{\ast}}^{2}(4-n)(1-n_{s})}\right]^{\frac{\beta_{2}}{2}},
\end{eqnarray}
where
\begin{equation}\nonumber
\alpha_{2}=\frac{2\kappa(\mu+2)^2(A^{\ast}f^{\ast})^2{\alpha_{0}}
^{\frac{4(f^{\ast}-1)}{f^{\ast}}}}{9\pi^2\alpha_{1}},
\quad\beta_{2}=\beta_{1}-\frac{4(1-f^{\ast})}{f^{\ast}}.
\end{equation}
During weak dissipative regime, we have calculated constraint on the
parameter $C_{\psi}$ using Eqs.(\ref{b}), (\ref{c}) and (\ref{d}) in
which $P_{R}=2.43\times10^{-9}$. It is proved that in anisotropic
model the dissipation coefficient $R$ evolutes during weak
dissipative regime for these constraints (Figures \textbf{1} and
\textbf{2}). The range of parameter $C_{\psi}$ for three different
values of $n$ and corresponding $f^{\ast}$ are given in Table
\textbf{1}. The parameter $f^{\ast}$ is obtained through
Eq.(\ref{a}) by fixing $N=70$ and $n_{s}=0.96$. Figures \textbf{3}
and \textbf{4} show the dependence of $r$ on $n_{s}$ for specific
values of the parameters. The left plot of Figure \textbf{3} (blue
curve) shows that warm anisotropic model is well fitted with recent
observations (WMAP9, Plank, BICEP2) for $C_{\psi}<10^{-4}$ during
weak intermediate regime. We have also computed an upper bound,
$C_{\psi}>10^{-6}$ that improves the compatibility of our model with
recent observations. Planck data places stronger bound on $r,~n_{s}$
as compared to WMAP9 and BICEP2. Figure \textbf{3} (right panel) is
plotted for $n=0$ and three different values of $C_{\psi}$ make the
model well supported by the WMAP9, Planck as well as BICEP2 for
$C_{\psi}<10^{-6}$. Similarly, for $n=-1$ (Figure \textbf{4}), our
model remains compatible with recent observations during
$10^{-12}<C_{\psi}<10^{-9}$. We have found that the presence of
anisotropic parameter $\mu$ leads to increase the values of
$C_{\psi}$ as compared to FRW \cite{29}. It is also noted that the
value of $C_{\psi}$ decreases with the increase of $n$.
\begin{table}
\caption{Constraints on $C_{\psi}$ for different values of $n$}
\vspace{0.5cm} \centering
\begin{small}
\begin{tabular}{|c|c|c|}
\hline\textbf{$n$}&\textbf{$f^{\ast}$}
&Constraint on \textbf{$C_{\psi}$}\\
\hline1&0.42&$10^{-6}<C_{\psi}<10^{-4}$\\
\hline0&0.37&$10^{-9}<C_{\psi}<10^{-6}$\\
\hline-1&0.34&$10^{-12}<C_{\psi}<10^{-9}$\\
\hline
\end{tabular}
\end{small}
\end{table}
\begin{figure}
\centering\epsfig{file=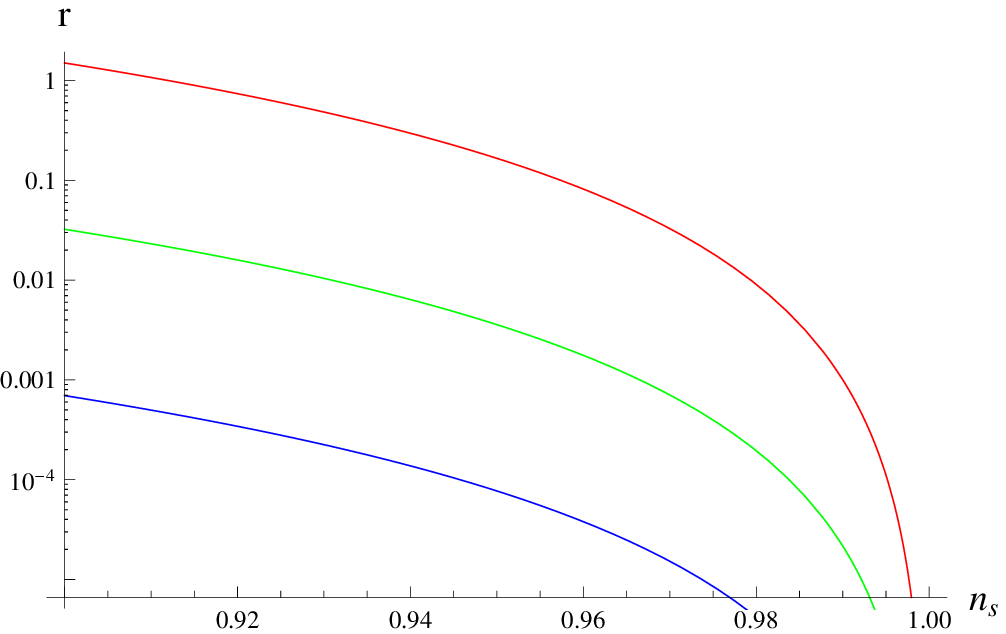,
width=0.50\linewidth}\epsfig{file=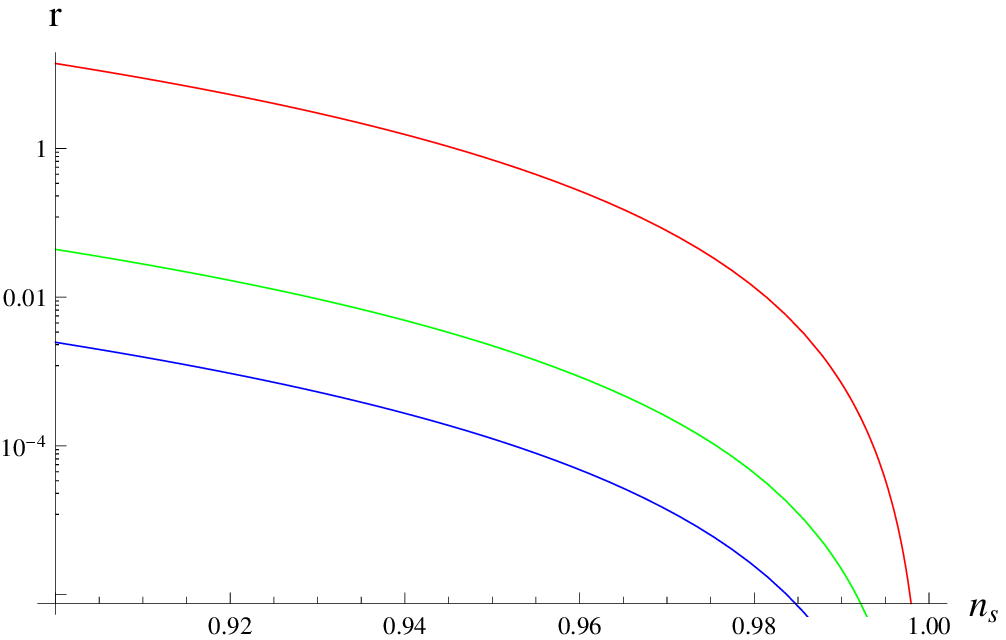,
width=0.50\linewidth}\caption{Plot of $r$ versus $n_{s}$: Left plot
for $f^{\ast}=0.42,~\mu\approx0.5,~n=1,~C_{\gamma}=70$,
$C_{\psi}=10^{-6}$ (red), $C_{\psi}=10^{-5}$ (green),
$C_{\psi}=10^{-4}$ (blue); (Right)
$f^{\ast}=0.37,~n=0,~C_{\psi}=10^{-9}$ (red), $C_{\psi}=10^{-7}$
(green), $C_{\psi}=10^{-6}$ (blue) during intermediate era.}
\end{figure}
\begin{figure}
\center\epsfig{file=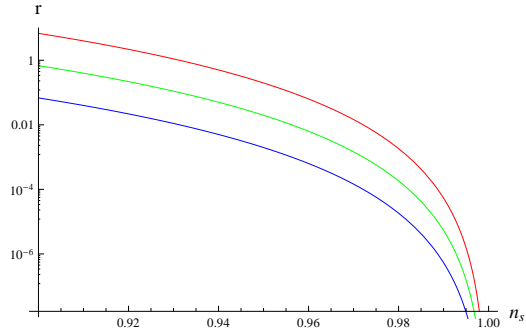, width=0.50\linewidth}\caption{Plot of
$r$ versus $n_{s}$ for
$f^{\ast}=0.34,~\mu\approx0.5,~n=-1,~C_{\gamma}=70$,
$C_{\psi}=10^{-12}$ (red), $C_{\psi}=10^{-11}$ (green),
$C_{\psi}=10^{-9}$ (blue).}
\end{figure}

\subsection{Logamediate Era}

During logamediate regime, the scale factor evolutes as \cite{17}
\begin{equation}\label{17}
Y(t)=Y_{0}\exp[A^{\ast}(\ln t)^{\lambda^{\ast}}].
\end{equation}
Using this value, inflaton takes the form
\begin{equation}\label{18}
\psi(t)-\psi(t_{0})=\left(\frac{2(1+2\mu)(A^{\ast}\lambda^{\ast})}
{\kappa(\mu+2)}\right)^{\frac{1}{2}}
\left(\frac{2}{\lambda^{\ast}+1}\right)(\ln
t)^{\frac{\lambda^{\ast}+1}{2}}.
\end{equation}
The cosmic time can be evaluated from the above equation as
\begin{equation}\nonumber
t=\exp\left[-B_{0}\psi^{\nu}\right],
\end{equation}
where
$B_{0}=\left(\frac{\kappa(\mu+2)}{2\nu(1+2\mu)(A^{\ast}\lambda^{\ast})}
\right)^{\frac{\nu}{2}},~\nu=\frac{2}{\lambda^{\ast}+1}$.
Substituting the value of $t$, we obtain
\begin{equation}\nonumber
H_{2}(\psi)=(A^{\ast}\lambda^{\ast}){B_{0}}^{(\lambda^{\ast}-1)}
\psi^{\nu(\lambda^{\ast}-1)}\exp\left[-B_{0}\psi^{\nu}\right],
\end{equation}
and $V(\psi)$ becomes
\begin{equation}\nonumber
V(\psi)=V_{0}\psi^{\alpha_{3}}\exp\left[-\beta_{3}\psi^{\nu}\right],
\end{equation}
where
$V_{0}=\left(\frac{1+2\mu}{\kappa}\right)(A^{\ast}\lambda^{\ast})^{2}
{B_{0}}^{2(\lambda^{\ast}-1)},
~\alpha_{3}=2\nu(\lambda^{\ast}-1),~\beta_{3}=2B_{0}.$ Here the
dissipation coefficient can be represented as
\begin{equation}\nonumber
\Gamma(\psi)=C_{\psi}^{\frac{4}{4-n}}\left[\frac{3(1+2\mu)}{2\kappa
C_{\gamma}(\mu+2)^2}\right]^{\frac{n}{4-n}}\psi^{\frac{4(1-n)}{4-n}}
\exp\left[\left(\frac{n}{n-4}\right){B_{0}}^{\nu}\psi^{\nu}\right].
\end{equation}
In weak dissipation regime, the decay rate
$R=\frac{\Gamma}{(\mu+2)H_{2}}<1$ leads to constrain $C_{\psi}$ as
given in Table \textbf{2}. Figure \textbf{5} shows that the range
$10^{-7}<C_{\psi}<10^{-3}$ for $n=1$ is compatible with weak
dissipative regime. It is found that the evolution of $R$ versus
$n_{s}$ for $n=0,-1$ also remains less than unity (graphs are not
shown).
\begin{table}
\caption{Constraints on $C_{\psi}$ for different values of $n$}
\vspace{0.5cm}\centering
\begin{small}
\begin{tabular}{|c|c|}
\hline\textbf{$n$}
&Constraint on \textbf{$C_{\psi}$}\\
\hline1&$10^{-7}<C_{\psi}<10^{-3}$\\
\hline0&$10^{-11}<C_{\psi}<10^{-7}$\\
\hline-1&$10^{-13}<C_{\psi}<10^{-10}$\\
\hline
\end{tabular}
\end{small}
\end{table}

In logamediate inflation, the slow-roll parameters are
\begin{eqnarray}\nonumber
\epsilon&=&\left(\frac{3}{\mu+2}\right)(A^{\ast}\lambda^{\ast})^{-1}
{B_{0}}^{-(\lambda^{\ast}-1)}\psi^{-\nu(\lambda^{\ast}-1)},\\\nonumber
\eta&=&\left(\frac{3}{\mu+2}\right)(A^{\ast}\lambda^{\ast}
{B_{0}}^{\lambda^{\ast}})^{-1}\psi^{-\nu\lambda^{\ast}}\left[2B_{0}\psi^{\nu}
-(\nu-1)\right].
\end{eqnarray}
The range of inflaton is
\begin{equation}\nonumber
\psi>\left[\left(\frac{\mu+2}{3}\right)(A^{\ast}\lambda^{\ast})
{B_{0}}^{(\lambda^{\ast}-1)}\right]
^{-\frac{1}{\nu(\lambda^{\ast}-1)}},
\end{equation}
and the initial inflaton at $\epsilon=1$ is
\begin{equation}\nonumber
\psi_{1}=\left[\left(\frac{\mu+2}{3}\right)(A^{\ast}\lambda^{\ast})
{B_{0}}^{(\lambda^{\ast}-1)}\right]
^{-\frac{1}{\nu(\lambda^{\ast}-1)}}.
\end{equation}
The warm model has the following number of e-folds
\begin{equation}\nonumber
N=\left(\frac{\mu+2}{3}\right)(A^{\ast}{B_{0}}^{\lambda^{\ast}})\left(\psi_{2}
^{\nu\lambda^{\ast}}-\psi_{1}^{\nu\lambda^{\ast}}\right).
\end{equation}
The inflaton dependent $P_{R}$ turns out to be
\begin{eqnarray}\nonumber
P_{R}(\psi)&=&\left(\frac{\kappa(\mu+2)^{4}}{6(1+2\mu)}\right)
\left[\frac{3(1+2\mu)C_{\phi}}{2\kappa
C_{\gamma}(\mu+2)^2}\right]^{\frac{1}{4-n}}(A^{\ast}\lambda^{\ast})^2
{B_{0}}^{2(\lambda^{\ast}-1)}
\psi^{(\alpha_{3}+\frac{1-n}{4-n})}\\\label{19}&\times&\exp
\left[-\left(\frac{5-n}{4-n} \right)B_{0}\psi^{\nu}\right].
\end{eqnarray}
Another inflaton is defined by $\psi_{2}$ as
\begin{equation}\nonumber
\psi_{2}=\left[\left(\frac{3}{\mu+2}\right){B_{0}}^{-\lambda^{\ast}}
\left[\frac{N}{A^{\ast}}+\left(\left(\frac{\mu+2}{3}\right)
(A^{\ast}\lambda^{\ast})^{\lambda^{\ast}}\right)^{-\frac{1}{\lambda^{\ast}-1}}
\right]\right]^{\frac{1}{\nu\lambda^{\ast}}}.
\end{equation}
Equation (\ref{19}) can also be written in terms of $N$ using
$\psi_{2}$ as
\begin{eqnarray}\nonumber
P_{R}(N)&=&\beta_{4}\left[\frac{N}{A^{\ast}}+\left(\left(\frac{\mu+2}{3}\right)
(A^{\ast}\lambda^{\ast})^{\lambda^{\ast}}\right)^{-\frac{1}{\lambda^{\ast}-1}}\right]^{\frac{1}
{\nu\lambda^{\ast}}(\alpha_{3}+\frac{1-n}{4-n})}\exp\left[\left(\frac{5-n}{n-4}\right)
\right.\\\label{20}&\times&\left.\left(\frac{3}{\mu+2}\right)^{\frac{1}{\lambda^{\ast}}}
\left[\frac{N}{A^{\ast}}+\left(\left(\frac{\mu+2}{3}\right)
(A^{\ast}\lambda^{\ast})^{\lambda^{\ast}}\right)^{-\frac{1}{\lambda^{\ast}-1}}
\right]^{\frac{1}{\lambda^{\ast}}}\right],
\end{eqnarray}
where
\begin{eqnarray}\nonumber
\beta_{4}&=&\left(\frac{\kappa(\mu+2)^4}{6(1+2\mu)}\right)
\left[\frac{3(1+2\mu)C_{\psi}}{2\kappa
C_{\gamma}(\mu+2)^{2}}\right]^{\frac{1}{4-n}}
\left(\frac{3}{\mu+2}\right)^{\frac{1}{\nu\lambda^{\ast}}
(\alpha_{3}+\frac{1-n}{4-n})}(A^{\ast}\lambda^{\ast})^{2}
\\\nonumber&\times&{B_{0}}^{2(\lambda^{\ast}-1)-\frac{1}{\nu}
(\alpha_{3}+\frac{1-n}{4-n})}.
\end{eqnarray}
\begin{figure}
\center\epsfig{file=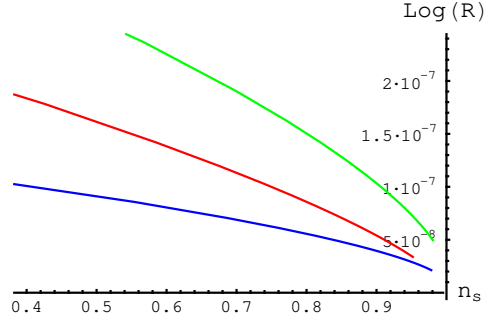, width=0.50\linewidth}\caption{Log$(R)$
versus $n_{s}$ for $\mu\approx0.5,~n=1,~C_{\gamma}=70,
~A^{\ast}=6.65\times10^{-3},~\lambda^{\ast}=6,~C_{\psi}=10^{-7}$
(red),
$A^{\ast}=8.43\times10^{-3},~\lambda^{\ast}=4.95,~C_{\psi}=10^{-5}$
(green)
$A^{\ast}=2.88\times10^{-2},~\lambda^{\ast}=4.36~,C_{\psi}=10^{-3}$
(blue).}
\end{figure}

The spectral index for the present model becomes
\begin{eqnarray}\nonumber
n_{s}-1&=&\left(\frac{3}{\mu+2}\right)\left[-\left(\frac{4(\lambda^{\ast}-1)}
{A^{\ast}\lambda^{\ast}}+\frac{(1-n)(\lambda^{\ast}+1)}{2(A^{\ast}\lambda^{\ast})(4-n)}\right)
{B_{0}}^{-\lambda^{\ast}}\psi^{-\nu\lambda^{\ast}}\right.\\\nonumber&+&\left.\left(\frac{5-n}{4-n}\right)
\frac{{B_{0}}^{(1-\lambda^{\ast})}
\psi^{\nu(1-\lambda^{\ast})}}{A^{\ast}\lambda^{\ast}}\right]\\\nonumber&=&-
\left[\frac{N}{A^{\ast}}+\left(\left(\frac{\mu+2}{3}\right)
(A^{\ast}\lambda^{\ast})^{\lambda^{\ast}}\right)^{-\frac{1}{\lambda^{\ast}-1}}\right]^{-1}
\left(\frac{(1-n)(\lambda^{\ast}+1)}{2(A^{\ast}\lambda^{\ast})(4-n)}
\right.\\\nonumber&+&\left.\frac{4(\lambda^{\ast}-1)}
{A^{\ast}\lambda^{\ast}}\right)
+\frac{1}{A^{\ast}\lambda^{\ast}}\left(\frac{5-n}{4-n}\right)
\left(\frac{3}{\mu+2}\right)^{\frac{1}{\lambda^{\ast}}+1}\\\label{f}&\times&\left[\frac{N}{A^{\ast}}
+\left(\left(\frac{\mu+2}{3}\right)
(A^{\ast}\lambda^{\ast})^{\lambda^{\ast}}\right)^{-\frac{1}{\lambda^{\ast}-1}}
\right]^{\frac{1-\lambda^{\ast}}{\lambda^{\ast}}}.
\end{eqnarray}
The tensor power spectrum is
\begin{equation}\nonumber
P_{T}(\psi)=\frac{2\kappa}{9\pi^2}
(\mu+2)^2(A^{\ast}\lambda^{\ast})^{2}{B_{0}}^{2(\lambda^{\ast}-1)}
\psi^{2\nu(\lambda^{\ast}-1)} \exp\left[-B_{0}\psi^{\nu}\right],
\end{equation}
which leads to the following $r$ along with Eq.(\ref{20})
\begin{eqnarray}\nonumber
r(\psi)&=&\frac{4(1+2\mu)}{3\pi^2(\mu+2)^2}\left[\frac{2\kappa
C_{\gamma}(\mu+2)^2}{3(1+2\mu)C_{\psi}}\right]^{\frac{1}{4-n}}\psi^{-\frac{1-n}{4-n}}
\exp\left[\left(\frac{5-n}{4-n}-2\right)B_{0}\psi^{\nu}\right]\\\nonumber&=&\beta_{5}
\left[\frac{N}{A^{\ast}}+\left(\left(\frac{\mu+2}{3}\right)
(A^{\ast}\lambda^{\ast})^{\lambda^{\ast}}\right)^{-\frac{1}{\lambda^{\ast}-1}}\right]
^{-\frac{(\lambda^{\ast}+1)(1-n)}
{2\lambda^{\ast}(4-n)}}\exp\left[\left(\frac{5-n}{4-n}-2\right)\right.
\\\label{g}&\times&\left.\left(\frac{3}{\mu+2}\right)
^{\frac{1}{\lambda^{\ast}}}\left[\frac{N}{A^{\ast}}+\left(\left(\frac{\mu+2}{3}\right)
(A^{\ast}\lambda^{\ast})^{\lambda^{\ast}}\right)^{-\frac{1}{\lambda^{\ast}-1}}
\right]^{\frac{1}{\lambda^{\ast}}}\right].
\end{eqnarray}
where
\begin{equation}\nonumber
\beta_{5}=\left(\frac{4(1+2\mu)}{3\pi^2(\mu+2)^2}\right)\left[\frac{2\kappa
C_{\gamma}(\mu+2)^{2}}{3(1+2\mu)C_{\psi}}\right]^{\frac{1}{4-n}}\left(\frac{\mu+2}{3}
\right)^{\frac{(\lambda^{\ast}+1)(1-n)}{2\lambda^{\ast}(4-n)}}
{B_{0}}^{\frac{(\lambda^{\ast}+1)(1-n)}{2(4-n)}}.
\end{equation}

\begin{table}
\caption{Constraints on $C_{\psi},~A^{\ast}$ and $\lambda^{\ast}$
for different values of $n$} \vspace{0.5cm}\centering
\begin{small}
\begin{tabular}{|c|c|c|c|}
\hline\textbf{$n$}
&\textbf{$C_{\psi}$}&\textbf{$\lambda^{\ast}$}&\textbf{$A^{\ast}$}\\
\hline1&$10^{-7}$&6.00&$6.65\times10^{-3}$\\
\hline1&$10^{-5}$&4.95&$8.43\times10^{-3}$\\
\hline1&$10^{-3}$&4.36&$2.88\times10^{-2}$\\
\hline0&$10^{-11}$&4.00&$1.79\times10^{-2}$\\
\hline0&$10^{-9}$&4.35&$1.16\times10^{-2}$\\
\hline0&$10^{-7}$&4.57&$4.53\times10^{-2}$\\
\hline-1&$10^{-13}$&4.85&$3.15\times10^{-2}$\\
\hline-1&$10^{-12}$&3.95&$1.99\times10^{-2}$\\
\hline-1&$10^{-10}$&3.25&$8.33\times10^{-2}$\\
\hline
\end{tabular}
\end{small}
\end{table}
\begin{figure}
\centering\epsfig{file=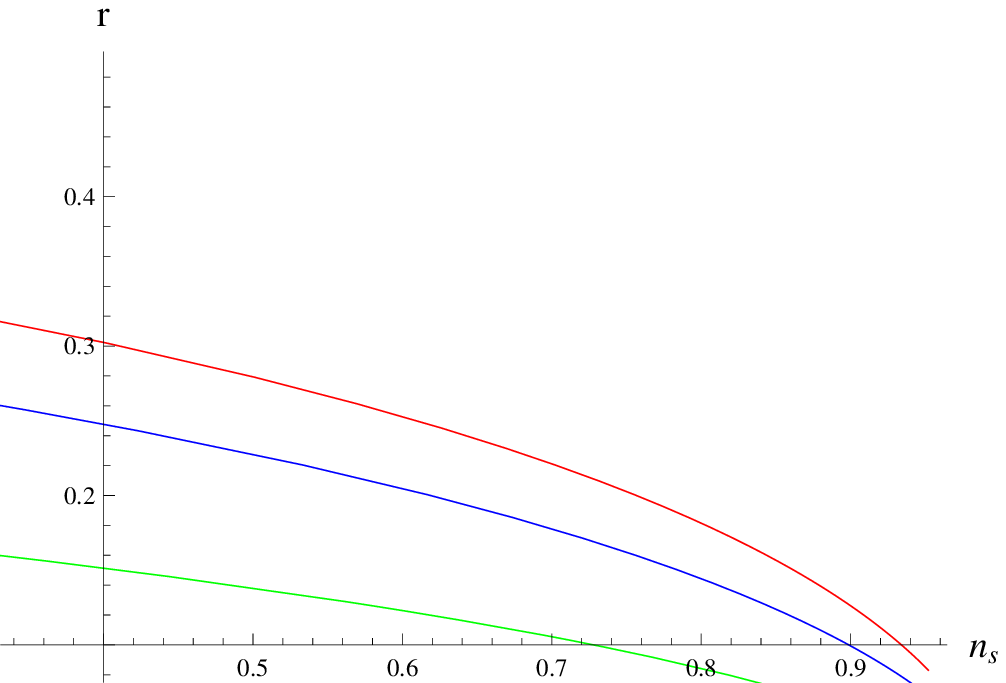,
width=0.50\linewidth}\epsfig{file=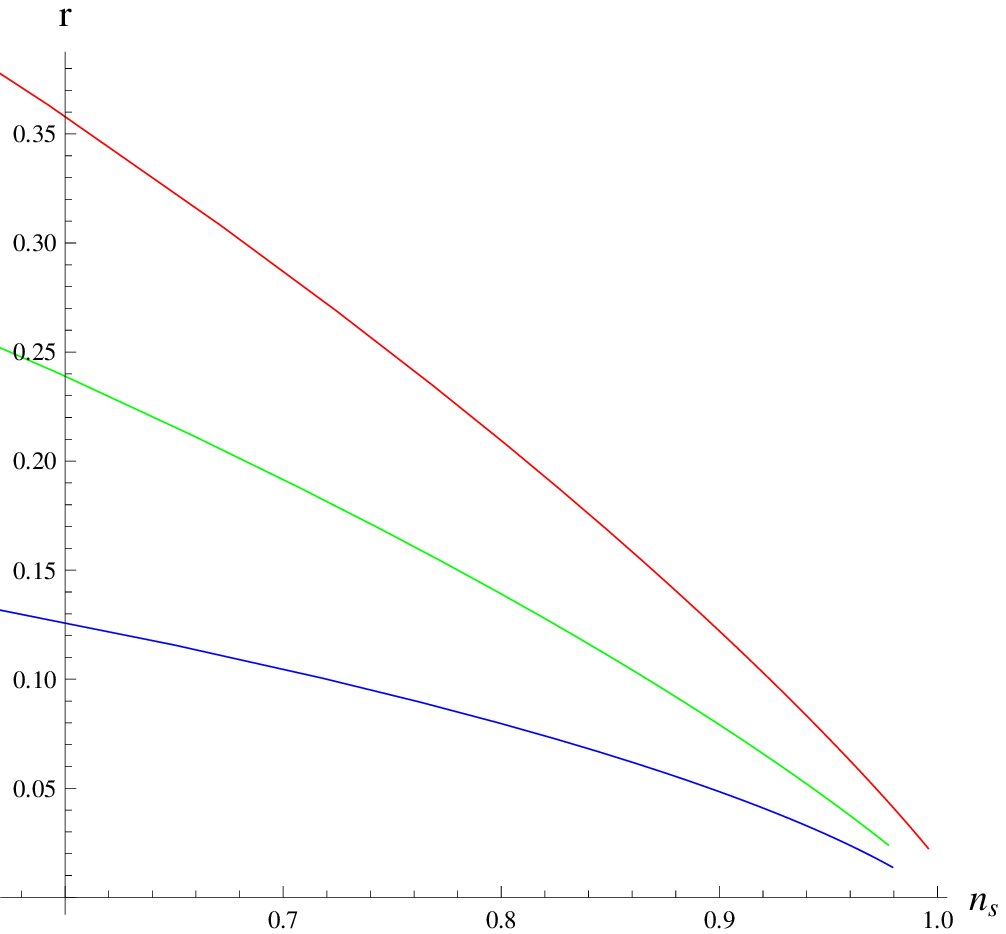,
width=0.45\linewidth}\caption{Plot of $r$ versus $n_{s}$: Left plot
for $\mu\approx0.5,~n=1,~C_{\gamma}=70$,
$A^{\ast}=6.65\times10^{-3},~\lambda^{\ast}=6,~C_{\psi}=10^{-7}$
(red),
$A^{\ast}=8.43\times10^{-3},~\lambda^{\ast}=4.95,~C_{\psi}=10^{-5}$
(green)
$A^{\ast}=2.88\times10^{-2},~\lambda^{\ast}=4.36~,C_{\psi}=10^{-3}$
(blue); (Right)
$A^{\ast}=1.79\times10^{-2},~n=0,~\lambda^{\ast}=4,~C_{\psi}=10^{-11}$
(red),
$A^{\ast}=1.16\times10^{-2},~\lambda^{\ast}=4.35,~C_{\psi}=10^{-9}$
(green),
$A^{\ast}=4.53\times10^{-2},~\lambda^{\ast}=4.57,~C_{\psi}=10^{-7}$
(blue) during logamediate era.}
\end{figure}
\begin{figure}
\center\epsfig{file=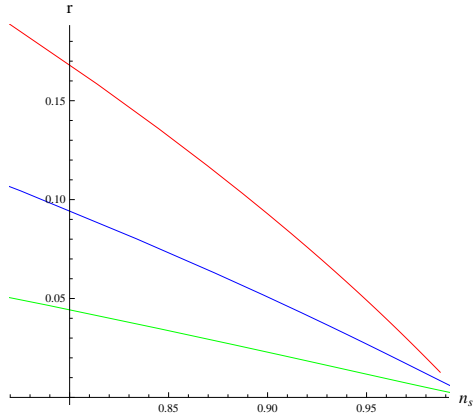, width=0.45\linewidth}\caption{Plot of
$r$ versus $n_{s}$ for $\mu\approx0.5,~n=-1,~C_{\gamma}=70$,
$A^{\ast}=3.15\times10^{-2},~\lambda^{\ast}=4.85,~C_{\psi}=10^{-13}$
(red),
$A^{\ast}=1.99\times10^{-2},~\lambda^{\ast}=3.95,~C_{\psi}=10^{-12}$
(green),
$A^{\ast}=8.33\times10^{-2},~\lambda^{\ast}=3.25,~C_{\psi}=10^{-10}$
(blue).}
\end{figure}
In order to constrain the physical parameters used in warm
logamediate model, we numerically solve Eqs.(\ref{20}) and (\ref{f})
for three different values of $n$. The values of the model
parameters $A^{\ast}$ and $\lambda^{\ast}$ for particular $C_{\psi}$
are picked up from the defined range are given in Table \textbf{3}.
Figures \textbf{6} and \textbf{7} show two-dimensional marginalized
constraints on the inflationary parameters $r$ and $n_{s}$ derived
from recent data. These graphs are plotted for three different
values of $A^{\ast},~\lambda^{\ast}$ and $C_{\psi}$. The left panel
of Figure \textbf{6} proves that our anisotropic warm inflationary
model is compatible with recent observations in the range
$10^{-7}<C_{\psi}<10^{-3}$ as $n_{s}=0.96$ generates very weak bound
$r<0.3$. Analogously, Figure \textbf{6} (right panel) verifies that
the range $10^{-11}<C_{\psi}<10^{-7}$ remains well consistent with
recent observations. Our anisotropic model satisfies the bounds of
WMAP9, Planck and BICEP2 for $C_{\psi}<10^{-10}$ (Figure
\textbf{7}).

One of the characteristics of warm inflation is $T>H_{2}$, where $T$
is the temperature of thermal background of the radiation. Further,
we assume that $T=T_{r}$, it is not a free parameter, we can
restrict $T_{r}>5.47\times10^{-5}$ using recent Planck data
$n_{s}=0.96\pm0.0073$ and an upper bound for $r<0.11$ at the pivot
point $k_{0}=0.002Mpc^{-1}$. We have checked the condition of warm
inflation ($T_{r}>H_2$) for our model. Figure \textbf{2} shows a
restriction on $T_{r}=T$, i.e., $T_{r}>H$.
\begin{figure} \center\epsfig{file=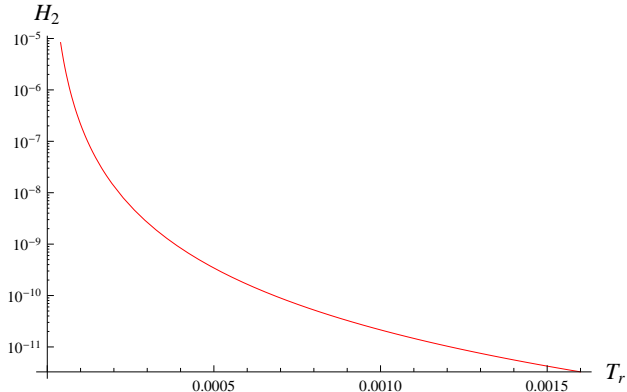,
width=0.60\linewidth}\caption{$H_2$ is plotted versus $T_r$.}
\end{figure}

\section{Concluding Remarks}

The idea of warm inflation is influenced by the friction term in the
inflaton equation of motion. The magnitude of the damping term
suggests the possibility that it could be the dominant effect
prolonging inflation. In this paper, we study the possible
realization of an expanding intermediate and logamediate scale
factors in weak dissipative regime and analyze how these two types
of inflation work with generalized form of the dissipation
coefficient $(\Gamma=C_{\psi}\frac{T^{n}}{\psi^{n-1}})$. To this
end, we have used the framework of homogeneous but anisotropic LRS
BI universe model which is asymptotically equivalent to the standard
FRW universe.

Recent data from the Planck satellite prove that large angle
anomalies represent real feature of the CMB map of the universe.
This result has a key importance since the small temperature
anisotropies and large angle anomalies may be caused by some unknown
mechanism or an anisotropic phase during the evolution of the
universe. This statement is particularly interesting because it
helps to modify the present model or to construct an alternative
model to decode the effects of the early universe on the present day
LSS without affecting the processes in the nucleosynthesis
\cite{30}. Scalar field leads to isotropisation, necessary for
compatibility with standard cosmological models at late times, as
well as inflation which accounts for structure formation.
Anisotropic model provides bouncing solutions that corresponds to
the late-time evolution which is driven to isotropy and spatial
flatness. Inflationary models has ability to make a transition form
early anisotropic phase to late-time isotropic evolution.

We have assumed that the universe is composed of standard scalar
field and radiation. Under slow-roll approximation, we have found
solutions of the first evolution equation (field equation) in weak
dissipative regime. During intermediate and logamediate eras, the
explicit expressions for inflaton $(\psi)$, corresponding effective
potential $(V(\psi))$ and rate of dissipation $(R)$ are calculated.
Moreover, we have evaluated perturbation parameters including
slow-roll parameters $(\epsilon,~\eta)$ to find the more general
conditions on the starting and ending conditions for the occurrence
of inflationary era, scalar and tensor power spectra
$(P_{R},~P_{T})$, scalar spectral index $(n_{s})$ and finally
observational parameter of interest, i.e., tensor to scalar ratio
$(r)$. In each case, we have constrained the model parameters
$(A^{\ast},~f^{\ast},~\lambda^{\ast},~C_{\psi})$ by WMAP9, Planck
and BICEP2 data for three particular values of $n=1,0,-1$.

In both regimes, we have proved that the values of the model
parameters given in Tables \textbf{1}-\textbf{3} lead to $R<1$ as
shown in Figures \textbf{1}, \textbf{2} and \textbf{5}. We conclude
that theses constraints make our warm anisotropic inflationary model
with generalized form of $\Gamma$ well supported by recent
observations. The trajectories of $r-n_{s}$ plotted in both regimes
are the verification of our results. The results of this paper
deviate from FRW universe in the following way. During weak
intermediate era, it is observed that due to the presence of
anisotropic parameter, $\mu$, constraints on the important model
parameter $C_{\psi}$ increase as compared to FRW universe. For
example, in anisotropic universe (LRS BI), $n=1$ leads to
$10^{-6}<C_{\psi}<10^{-4}$ while for isotropic universe (FRW), this
is $10^{-9}<C_{\psi}<10^{-6}$. Further, it is observed that the
value of $C_{\psi}$ decreases with the increase of $n$ due to
anisotropic parameter. Since $C_{\psi}$ acts as coupling parameter,
so its decreasing nature is well consistent with observations, i.e.,
coupling between two constituents of the universe must be decreasing
with the cosmic evolution. This result has opposite effect as
compared to isotropic universe. Similarly, in logamediate regime, we
are successful in constraining $C_{\psi}<10^{-10}$ that leads to
compatibility of the model with WMAP9, Planck as well as BICEP2. It
is observed that the compatibility range of $C_{\psi}$ is less than
FRW universe.

It is also found that compatibility of the model disturbs for too
large values of the anisotropic parameter $\mu>10^{3}$. The case
$n=3$ where ($\Gamma\propto \frac{T^{3}}{\phi^2}$) is discussed for
strong dissipation regime during intermediate and logamediate
regimes \cite{23}. It is worth mentioning here that all the results
reduce to the isotropic universe for $\mu=1$ \cite{29} and $n=3$
leads to \cite{33}. The work in strong dissipative regime and
interpolation between weak and strong regimes is under process.

\vspace{0.5cm}

{\bf Acknowledgment}

\vspace{0.5cm}

We would like to thank the Higher Education Commission, Islamabad,
Pakistan for its financial support through the Indigenous Ph.D.
Fellowship for 5000 Scholars Phase-II, Batch-I.

\end{document}